\documentclass[preprint]{aastex6}
\shorttitle{RRATs with LWA1}
\shortauthors{Taylor et al.}

\usepackage{graphicx}	
\usepackage{amsmath}	
\usepackage{amssymb}	

\begin{document}
\label{firstpage}


\title{Observations of Rotating Radio Transients with the First Station of the Long Wavelength Array }

\author{G.B. Taylor\altaffilmark{1}}
\affil{Department of Physics and Astronomy, University of New
Mexico, Albuquerque NM, 87131, USA}
\altaffiltext{1}{Greg Taylor is also an Adjunct Astronomer at the National Radio Astronomy Observatory}

\author{K. Stovall}
\affil{Department of Physics and Astronomy, University of New
Mexico, Albuquerque NM, 87131, USA}

\author{M. McCrackan\altaffilmark{2}}
\affil{Department of Physics and Astronomy, University of New
Mexico, Albuquerque NM, 87131, USA}
\altaffiltext{2}{Now at Department of Astronomy, University of Massachusetts, Amherst, MA 01003, USA}

\author{M.A. McLaughlin and R. Miller}
\affil{Department of Physics and Astronomy, West Virginia University, Morgantown, WV 26506-6315, USA}

\author{C. Karako-Argaman}
\affil{Department of Physics, McGill University, 3600 University Street, Montreal, QC H3A 2T8, Canada}

\and

\author{J. Dowell
and F.K. Schinzel\altaffilmark{3}}
\affil{Department of Physics and Astronomy, University of New
Mexico, Albuquerque NM, 87131, USA}
\altaffiltext{3}{Now at the National Radio Astronomy Observatory, Socorro NM, 87801, USA}

\begin{abstract}

Rotating Radio Transients (RRATs) are a subclass of pulsars first
identified in 2006 that are detected only in searches for single
pulses and not through their time averaged emission.  Here, we present
the results of observations  of 19 RRATs using the first station
of the Long Wavelength Array (LWA1) at frequencies between 30 MHz
and 88 MHz.  The RRATs observed here were first detected in higher
frequency pulsar surveys.  Of the 19
RRATs observed, 2 sources were detected and their dispersion measures
(DMs), periods, pulse profiles, and flux densities are reported
and compared to previous higher frequency measurements.  We find a low
detection rate (11\%), which could be a combination of the lower 
sensitivity of LWA1 compared to the higher frequency telescopes, 
and the result of scattering 
by the interstellar medium (ISM) or a spectral turnover.

\end{abstract}



\section{Introduction}

Rotating Radio Transients (RRATs) are a subclass of pulsars whose
emission is only observed intermittently, with only single pulses
being detected.  These single pulses can be separated from one another
on timescales ranging from minutes to hours.  In general RRATs are not
detected through periodicity searches as is the case with other
pulsars \citep{mcl06}.  Therefore, single pulse searches are needed to
identify them. However, the ability to detect RRATs depends upon the
instrument, the frequency and the length of the observations, and some
RRATs may even be detectable as normal pulsars in periodicity searches
for some combinations of parameters.

RRATs were first discovered in a re-analysis of data from the 1.4 GHz
Parkes Multibeam Pulsar Survey (PMPS), which identified 11 RRATs with
single pulses with pulse widths ranging from 2 ms to 30 ms, peak flux
densities between 0.1 Jy and 3.6 Jy, and detected pulses ranging from
only 4 up to 229 \citep{man01,mcl06}.  More than 100 RRATs have been
discovered thus far with most being found in the blind pulsar surveys
carried out by major radio observatories\footnote{\noindent See the
  RRATalog at http://astro.phys.wvu.edu/rratalog/}.  Some notable
surveys include the Green Bank Telescope's 350 MHz Drift Scan Survey
and its Northern Celestial Cap Pulsar Survey \citep{lyn13,kar15}, the
High Time Resolution Universe Survey \citep{sbs11}, Arecibo's 327 MHz
Drift-Scan Pulsar Survey \citep{den13}, as well as in subsequent
re-analyses of the PMPS \citep{kea10b,kea11}.

The mechanism behind the transient pulsing of RRATs has not yet been
determined, with many models having been put forth.  The inferred
periods of the still small fraction of RRATs that have measured
periods are generally longer than the typical periods of canonical
pulsars, with Karako-Argman et al. tentatively finding some evidence
that pulsars with periods longer than 200 ms are more likely to be 
detectable only in single-pulse searches than pulsars with shorter
periods \citep{mcl06,kea10b,kar15}.  It has been suggested that
RRATs are old pulsars near the end of their pulsing lifetime and turn
off in a sporadic manner \citep{zha07}.  Others suggest RRATs are
pulsars whose emission is regulated by a surrounding debris field
\citep{cor08}, or that they are fundamentally similar to nulling
pulsars \citep{sbs10}.  Another proposed model is that RRATs are
pulsars which emit single bright pulses in addition to a distribution
of weaker single pulses \citep{wel06}.

If RRATs are actually a distinct type 
of pulsar that forms independently of regular pulsars, they should be counted
in addition to the number of other pulsars.  Given the need for the
birth rates of the various populations of neutron stars to add
up to the supernova rate in the Galaxy, the presence of RRATs would
require a substantial increase in that supernova rate.  Indeed, the
maximum supernova rate was estimated by Keane et al. (2010) to be 
3 century$^{-1}$,
whereas the neutron star birthrate required to account for RRATs alone as a
distinct class was estimated to be 5.8 century$^{-1}$.

Using the first station of the LWA (LWA1), we carried out observations
of 19 RRATs at frequencies between 30 MHz and 88 MHz.  This is the
first time that RRATs have been studied at such low frequencies, and
their properties, including burst rates, flux densities, pulse shapes,
and spectral indices are consequently all unknown.  The LWA1 possesses
both high time and frequency resolution, thereby allowing the pulse
properties and dispersion measures (DMs) of detectable pulses to be
determined very precisely.  For detected RRATs, the LWA1 has more
available observing time than many other instruments, so timing
observations can be carried out readily.

We describe two observing campaigns and the data reduction in \S2, 
followed by the details of the two detected RRATs individually in
\S3 and the consider the implications in \S4, with concluding 
remarks in \S5.

\section{Observations}

The observations reported here were taken with the first station of
the Long Wavelength Array (LWA1; Taylor et al. 2012).   The observatory consists of
two stations, LWA1 and LWA-SV.  For this work we used LWA1, which is
co-located with the Jansky Very Large Array.  This station utilizes
256 stands of orthogonal dipoles and can observe at any frequency
between 10 MHz and 88 MHz with a bandwidth of 16 MHz.  Four
independent steerable beams have two frequency tunings each and a
beamwidth of 2.2 degrees at 74 MHz.   With its
high time and frequency resolution, the LWA is particularly well
suited to observations of time varying phenomena such as pulsars and
RRATs.  More than 60 regular pulsars have been re-detected with the LWA
with 4 of these being millisecond pulsars
\citep{sto15}.  Given the observatory's low frequency
range, where pulses will experience scattering and dispersion effects
to a much greater degree than in higher frequency observations, LWA1
can make very accurate studies of the time-varying properties of the 
interstellar medium (ISM) with uncertainties for the DM values 
in the range of 0.001 pc cm$^{-3}$.

Two different observing campaigns of RRATs were carried out with the
LWA1, with the first taking place from mid-2013 to early 2014 and the
second campaign taking place from August 2014 to March 2015.  In the
first observing session (LWA project code LM002) we selected 10 RRATs
with DMs less than 50 pc cm$^{-3}$, as the scattering timescale
given by~\cite{bhat04} at 64 MHz is over 100 milliseconds at this DM,
much larger than the typical width of a RRAT's pulse.   
These RRATs had also been confirmed
in multiple observations with different telescopes.  A single LWA1 
beam was utilized, which provided an effective central frequency of
72 MHz and a bandwidth of 32 MHz. The bandwidth of the individual 
channels used for analysis was 4.8 kHz.
Each observation was 2 hours in
length and any detections were followed up with additional
observations.  

The second observing campaign (LWA projects DM001 and LM003) selected
an additional 9 RRATs to be observed along with the 10 from the first
observing run with less stringent constraints on their DMs.  All of
these additional targets have DMs less than 100 pc cm$^{-3}$. While
the first observing run selected RRATs without previous timing
solutions, many of those in the second observing run have measured
periods allowing for comparisons with higher frequency results.  The
nature of the observations were changed to more resemble the style
utilized for the LWA1's already successful pulsar observations.  Two
of the beams were used in the split filter setting with central
frequencies of 35.1 MHz, 49.8 MHz, 64.5 MHz, and 79.2 MHz and an
effective bandwidth after combination of 60 MHz.  Each RRAT was
observed with 1 hour observations and any detections were followed up
with additional observations.  All observations of this run were
scheduled at night across transit to limit the level of radio
frequency interference (RFI).  All 19 RRATs targeted in either
campaign are listed in Table~1 along with their measured DMs and
periods from discovery, follow up, or from our observations.  For the
objects that we detect we also provide the low frequency burst rate.

\begin{table*}
\begin{center}
\footnotesize
\def\arraystretch{0.5}
Table 1. RRATs Observed with LWA1 \\
\begin{tabular}{lllllllll}
\hline 
\hline
\vspace{-0.1cm}\\
Source & DM & Period & Rate & Ref. & Freq. & BW & Time & Date$^\dagger$\\ 
 & (pc cm$^{-3}$) & (s) & (hr$^{-1}$) & & (MHz) & (MHz) & (hr) & (YY-MM-DD) \\
\hline 
\hline
J0054+69 & 90.3 & $-$ & & 1 & 57 & 60 & 1 & 14-11-08 \\
J0054+66 & 14.554 & 1.390 & 0.7 & 2,3 & 57 & 60 & 1 & 15-01-17 \\
 & &  & & &  57 & 60 & 3 & 14-12-28, 23, 20 \\
 & &  & & & 57 & 60 & 2 & 14-12-18, 17 \\
J0103+54 & 55.605 & 0.354 & & 1 &  57 & 60 & 1 & 14-11-26 \\ 
 &  &  & & & 57 & 60 & 1 & 14-10-05 \\ 
J0201+7005 & 19.998 & 1.349 &  & 1 & 72 & 32 & 2 & 13-10-17 \\
 &  & & & & 72 & 32 & 2 & 14-01-11 \\
 & &  & & &  57 & 60 & 2 & 14-12-19, 14-10-13 \\
 & &  & & &  57 & 60 & 2 & 14-11-17, 13 \\
 & &  & & &  57 & 60 & 3 & 14-09-29, 24, 21 \\
J0332+79 & 16.67 & 2.056 & & 1 & 72 & 32 & 2 & 13-10-17 \\
 & &  & & &  57 & 60 & 1 & 15-01-25 \\
 & &  & & &  57 & 60 & 1 & 14-10-06 \\
J0447$-$04 & 29.83 & 1.188  & & 1 & 72 & 32 & 2 & 13-10-17 \\
 & &  & & &  57 & 60 & 1 & 15-01-24 \\
 & &  & & &  57 & 60 & 1 & 14-11-10 \\
J0628+09 & 88 & 1.241 & & 4 &  57 & 60 & 1 & 14-11-09 \\
J0957$-$06 & 26.95 & 1.724  & & 1 & 72 & 32 & 2 & 14-01-08 \\
 & &  & & &  57 & 60 & 1 & 15-02-06 \\
 & &  & & &  57 & 60 & 1 & 14-12-27 \\
J1439+76 & 22.29 & 0.948 & & 1 & 72 & 32 & 2 & 14-01-08 \\
 & &  & & &  57 & 60 & 1 & 15-02-05 \\
 & &  & & &  57 & 60 & 1 & 15-01-19 \\
J1538+2345 & 14.909 & 3.449  & & 1 & 72 & 32 & 2 & 14-01-08 \\
 & &  & & &   57 & 60 & 1 & 15-02-27 \\
 & &  & & &  57 & 60 & 1 & 15-02-07 \\
 & &  & & &  57 & 60 & 1 & 15-01-16 \\
J1611$-$01 & 27.21 & 1.297 & & 1 &  72 & 32 & 2 & 14-01-08 \\
 &  &  & & &  57 & 60 & 1 & 15-07-17 \\
J1610$-$17 & 52 & $-$ & & 5 &  57 & 60 & 1 & 15-02-17 \\
 &  &  & & &  57 & 60 & 1 & 15-01-14 \\
J1623$-$0841 & 60.42 & 0.503 & & 6 & 57 & 60 & 1 & 15-02-18 \\
 & &  & &  & 57 & 60 & 1 & 15-01-15 \\
J1705$-$04 & 42.951 & 0.237 & & 1 & 72 & 32 & 2 & 14-01-08 \\
 &  &  & & & 57 & 60 & 1 & 15-07-18 \\
J1753$-$12 & 73 & 0.405 & & 5 & 57 & 60 & 1 & 15-02-19 \\
J1850+15 & 24 & 1.383 & &  5 & 57 & 60 & 1 & 14-12-21 \\
J1944$-$10 & 31.01 & 0.409 & & 1 & 72 & 32 & 2 & 13-07-23 \\
 & &  & & &  57 & 60 & 1 & 15-07-16 \\
J2225+35 & 51 & 0.94 & & 7 &  57 & 60 & 1 & 14-10-18 \\
 &  &  & & & 57 & 60 & 1 & 14-09-19 \\
J2325$-$0530 & 14.963 & 0.869 & 21  & 1,3 & 72 & 32 & 2 & 13-07-23 \\
 & &  & & & 72 & 32 & 2 & 14-01-11 \\
 & &  & & & 57 & 60 & 2 & 14-12-19, 14-11-14 \\
 & &  & & & 57 & 60 & 3 & 14-10-31, 26, 23 \\
 & &  & & & 57 & 60 & 3 & 14-10-21, 20, 19 \\
\hline
\multicolumn{9}{l}{References: (1) Karako-Argaman et al. 2015 (2) Hessels et al. 2008 (3) this work (4) Deneva et al. 2009} \\
\multicolumn{9}{l}{(5) Burke-Spolaor \& Bailes 2010 (6) Boyles et al. 2013 (7) Shitov et al. 2009} \\
\multicolumn{9}{l}{$^\dagger$ Epochs in YY-MM-DD format with additional entries corresponding to days in the same month.} \\
\end{tabular}
\end{center}
\end{table*}

The data reduction was carried out at the LWA's User Computing
Facility (LWAUCF), a computing cluster consisting of 6 nodes located at
the nearby VLA control building, with the PRESTO software
package\footnote{http://www.cv.nrao.edu/$\sim$sransom/presto} being used for the analysis \citep{ran01}.  
The raw data from each frequency tuning were first converted
to fits format using {\it write2psrfits.py} 
and the resulting files combined
to yield a single fits file for all frequency tunings
(see Dowell et al. 
2012\nocite{dow12}).  This
combination of the beams provides an effective increase in bandwidth,
thereby increasing the signal-to-noise ratio (S/N) of the pulses.
RFI was then flagged utilizing {\it rfifind} and the data were 
dedispersed into 128
subbands at trial DMs spanning 10 pc cm$^{-3}$ with a step size 
of 0.001 pc cm$^{-3}$ centered on the DMs of their
discovery detections. This large DM range was searched in order to ensure any real pulses were not missed due to a larger error than that which is reported in the discovery detection and to determine the false-positive detection rate resulting from any remaining RFI.   Due to the low observing frequency of these
observations, the dispersion of the pulses is significant even for low
DMs, so a small step size is required.
This results in a smearing due to an incorrect DM of roughly 1 millisecond
across the band.  The
results were searched using both single pulse search and
periodicity search techniques using the routines {\it single\_pulse\_search.py} and 
{\it prepfold} respectively.

Again, as a result of the $\nu^{-4.4}$ frequency dependence of
  scattering effects of the ISM, pulses in the LWA1's frequency band
  are scattered by a factor of ~2000 times more than a pulse at 350
  MHz \citep{cor02}.  It is possible that the pulses from the RRATs
might be too scattered to be detected by the standard match
filtering. For this reason a modified version of {\it single\_pulse\_search.py}, which performs a matched-filter search using Equation 3 from~\cite{kar12} with
$\mathrm{\tau_{d}}$ values ranging from 2 to 42 ms, was
also utilized to improve the chance of detection of some of the higher
DM RRATs that were observed.  This technique has also been successfully
applied to detect giant pulses from the Crab pulsar as described 
by \cite{tar16}.

To estimate the flux density of the detected pulses we employed the
radiometer equation which relates the signal-to-noise to the
integration time, bandwidth, and system equivalent flux density (SEFD)
of the telescope. The SEFD of the LWA1 varies with observing
frequency, projected area of the array as a result of the source's
elevation and the proximity to bright sources such as the Galactic
Plane.  We used the results of \cite{sch14} to estimate the response
 of the LWA1 at varying zenith angles, and from that derive the flux
densities for measured pulses.  The uncertainty we conservatively
assigned to this method is $\pm$50\% \citep{sto15}.

\section{Results}

Of the 19 RRATs observed with the LWA, just 2 confirmed detections have
been made, with one from the 2013 -- 2014 observing run and one
from the 2014 -- 2015 observing run.  The RRATs that
were detected are J0054+66 and J2325$-$0530, both of which were
originally discovered with the Green Bank Telescope (GBT) in its 350 MHz sky
surveys.  These will be addressed individually below.  

\subsection{J0054+66}

\begin{figure*}
\begin{center}
\includegraphics[width=5.0in,angle=0]{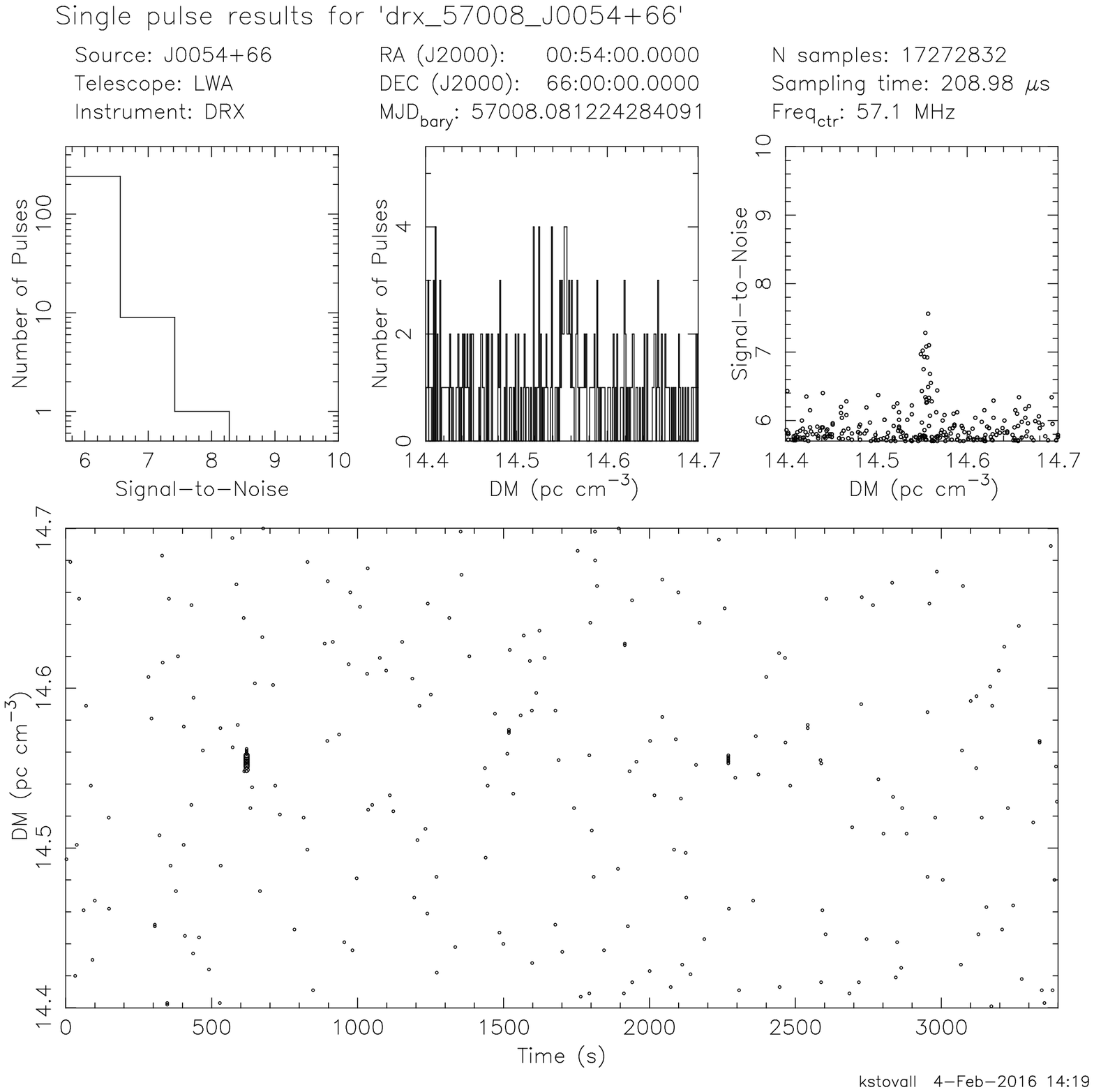}
\end{center}
\vspace{-0.5cm}
\caption{\label{fig:J0054} 
LWA1 detection of J0054+66 showing two pulses detected
with the point size scaling with S/N on the bottom, 
distribution of the number of pulses (over a range of DM) with 
S/N at top left, the distribution of the number of pulses with DM in the middle,
and the distribution of S/N with DM at top right.  
}
\end{figure*}

This RRAT was one of the 8 new RRATs selected for observing in the
second observing session, and was discovered with the GBT in its
Survey of the Northern Galactic plane for Radio Pulsars and Transients
at 350 MHz.  Its initial GBT detection DM was $\sim$ 14.5 pc
cm$^{-3}$ with no reported error, with a burst rate ranging from
$\sim$4 to 40 hr$^{-1}$, 
and a period of 
1.390 s \citep{hes08}.  The pulse profile from the GBT 
had a width on the order of 10 ms.
A total of 6 observations of 1 hour with the LWA1
have been carried out on this RRAT, with 3 yielding detections and the
others no observable emission.  To be classified as a detection we
required a minimum S/N of 6, and that the S/N vs DM peaked at a 
reasonable value for the DM.  The flux density of the faintest
detectable pulse for the combined 60 MHz band was 
$\sim$50 Jy.  The number of detectable pulses in
each of the observations varied, with two of the
combined bandwidth (60 MHz) observations yielding only one detected 
pulse per observation and one detecting 2 pulses as is illustrated in 
Fig.~\ref{fig:J0054}.  We estimate the burst rate from these observations
to be 0.7 hr$^{-1}$. 

\begin{figure}
\begin{center}
\includegraphics[width=3.5in,angle=0]{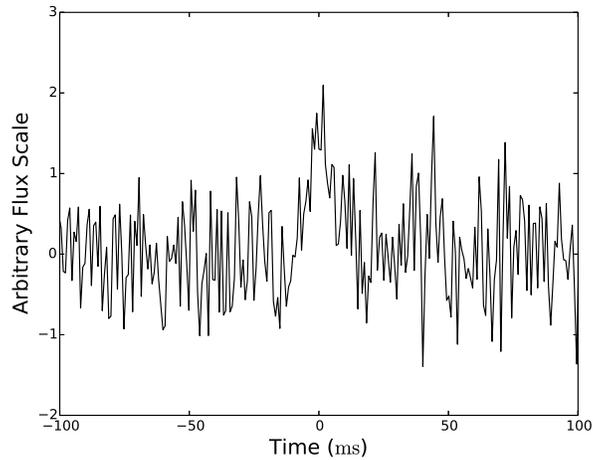}
\end{center}
\vspace{-0.5cm}
\caption{\label{fig:J0054p} 
The pulse profile of a single bright pulse from J0054+66 at a center frequency of 57 MHz with
a bandwidth of 60 MHz.
}
\end{figure}

The detections at the other individual frequencies without the
combined beam also vary in the number of pulses detected, with some
finding none, whereas the 64.5 MHz tuning of the same observation that
detected 2 pulses in the combined bandwidth also detected 2 pulses within
the hour long observation.  The pulse profile for a single bright pulse
observed with the LWA1 is shown in Fig.~\ref{fig:J0054p}.

The DM measured with LWA1 did not change significantly from
observation to observation or pulse to pulse in those observations
that had multiple detections, with a DM of 14.554 $\pm$ 0.007 pc
cm$^{-3}$.  To estimate the error we fit the signal-to-noise vs DM
curve to Eq.~12 from \cite{cor03} and took the error to correspond to
the range in which values are 1$\sigma$ below the peak. However,
we note that significant profile evolution, whether it is intrinsic or
due to scattering, would bias this DM value~\citep[e.g.][]{has12}. The average
S/N, mean flux and pulse width for J0054+66 are listed in
Table~2.  The signal-to-noise in the individual tunings was not
sufficient to allow measurements across the band.  The variations in
S/N are likely intrinsic to the RRAT and perhaps modulated
  by the ionosphere and/or ISM.  The elevation angle of the RRAT was
the same for all the observations, meaning that the LWA1's sensitivity
should have remained nearly constant for each detection.  The
variations in S/N and pulse widths result in a derived peak flux
density that is also variable, with values between about 100 Jy and
160 Jy for the combined bandwidth.


PSR J0054+66 was not detected in any periodicity searches folded
  at the period known from the GBT observations with the LWA1.  An
additional 2 hour observation was performed with the intention of
improving the detectability in a periodicity search and to get a
better estimate on the pulse rate, but no single pulse or periodic
emission were detected in this longer session. 
However, the period derived
  from the lowest common denominator method is consisitent with the
  period reported from the GBT detection.

\begin{table}
\begin{center}
Table 2. Pulse Properties for J0054+66 \\
\begin{tabular}{lllllll}
\hline 
\hline
\vspace{-0.1cm}\\
Frequency & FWHM & Peak Flux & Mean Flux& Max. S/N \\
 (MHz) & (ms) & (Jy) & (Jy) & \\
\hline 
\hline
57 & 11 $\pm$ 2 & 70 $\pm$ 35 & 0.6 $\pm$ 0.3 & 8.3 \\
\hline
\\
\end{tabular}
\end{center}
\end{table}

\subsection{J2325$-$0530}

\begin{figure*}
\begin{center}
\includegraphics[width=5.0in,angle=0]{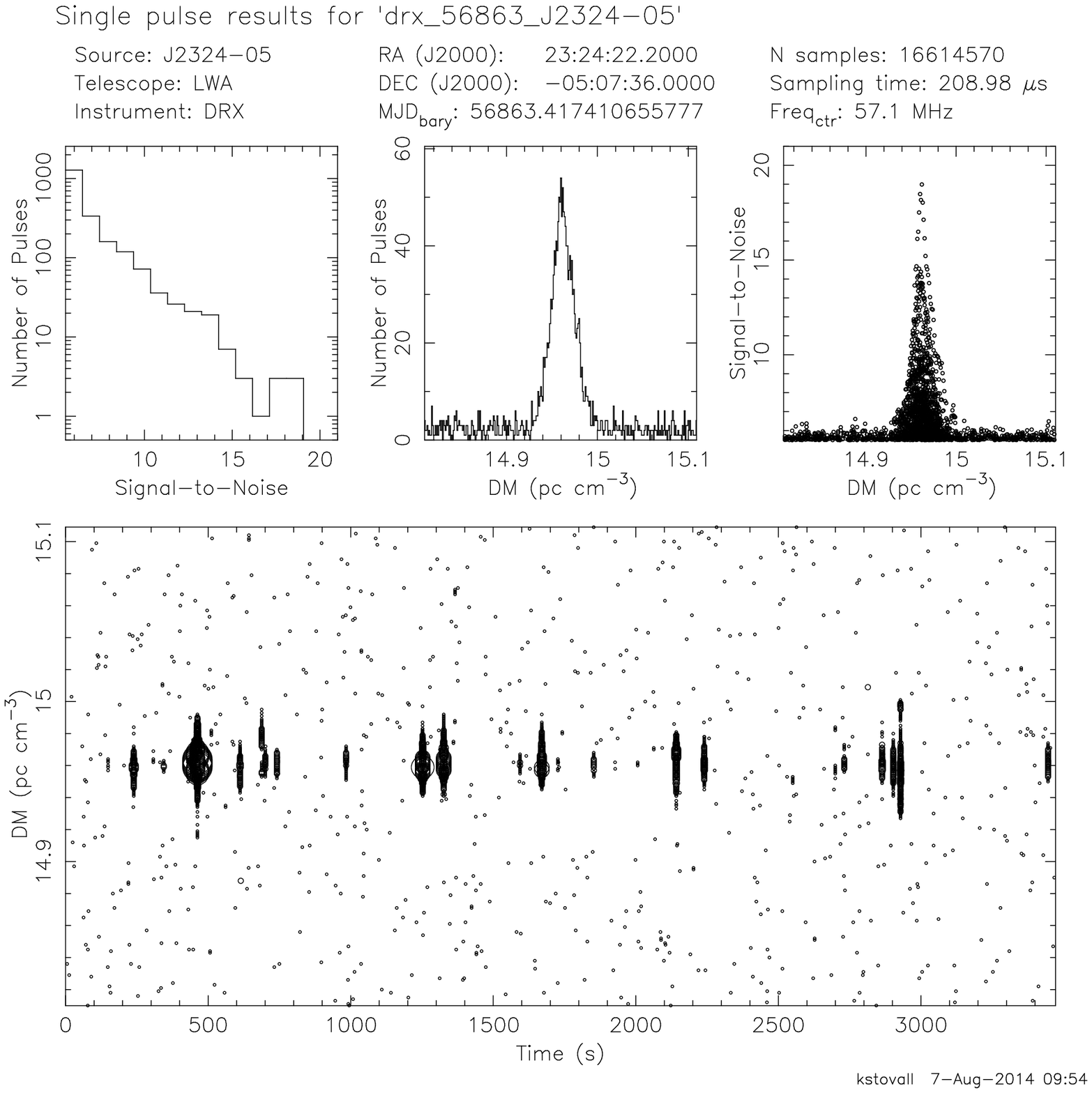}
\end{center}
\vspace{-0.5cm}
\caption{\label{fig:J2324} 
LWA1 detection of J2325$-$0530 showing pulse detection 
with the point size scaling with { S/N} on the bottom, 
distribution of the number of pulses with { S/N} at top left, 
the distribution of the number of pulses with DM in the middle,
and the distribution of { S/N} with DM at top right.  
}
\end{figure*}

\begin{figure}
\begin{center}
\includegraphics[width=3.5in,angle=0]{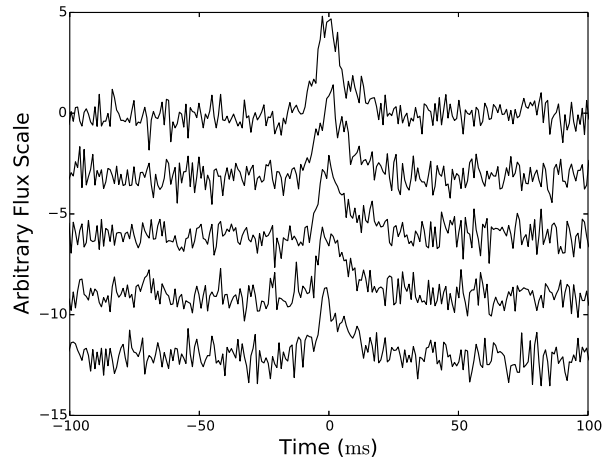}
\end{center}
\vspace{-0.5cm}
\caption{\label{fig:J2324p} 
LWA1 profiles at a center frequency of 57 MHz and with 60 MHz bandwidth
for five strong pulses from J2325$-$0530 showing a scattering tail.  
}
\end{figure}

PSR J2325$-$0530 was the first RRAT to be detected with the LWA and has
been detected in both observing campaigns.  Due to the wider bandwidth
of the second campaign, the S/N and burst rate are much higher than in the first observing run, as is shown
in Fig.~\ref{fig:J2324}.  This RRAT was originally found in the 2007 GBT's 
350 MHz Drift Scan Survey described previously, with a DM
of 14.966 $\pm$ 0.007 pc cm$^{-3}$ at a discovery S/N of 13.  
The burst rate in the 350 MHz GBT and the 110 MHz LOFAR observations were 46 $\pm$ 9 hr$^{-1}$ and 52 $\pm$ 8 hr$^{-1}$, respectively \citep{kar15}.  In comparison, the LWA1's
detection DM was 14.963 $\pm$ 0.006 pc cm$^{-3}$, with a peak {S/N} of 19.
For the faintest detectable pulses using the combined bandwidth we used a
S/N of 6 corresponding to a flux density of $\sim$60 Jy.
The pulse rate was also much higher than for the detection of J0054+66, with
the combined bandwidth observations yielding a rate of 21 hr$^{-1}$, compared to
the first observing run's 12 hr$^{-1}$.  Individual pulses have
similar shapes with a scattering tail out to $\sim$30 milliseconds at 35.1 MHz
as shown in Fig.~\ref{fig:J2324p}.

\begin{figure}
\begin{center}
\includegraphics[width=3.5in,angle=0]{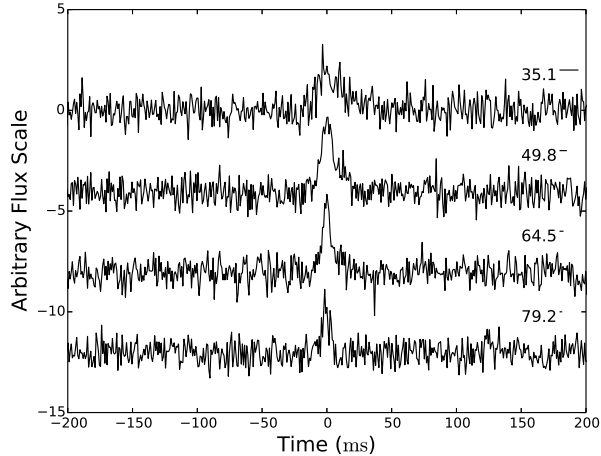}
\end{center}
\vspace{-0.5cm}
\caption{\label{fig:pfreq} 
LWA1 profiles for a single pulse from J2325$-$0530 shown at 
four different frequencies. The line next to the frequency label
indicates the timescale of the interchannel smearing due to dispersion.
}
\end{figure}

Pulse widths for a typical pulse in the individual frequency 
tunings (Fig.~\ref{fig:pfreq}) were found
and are listed along with their {S/N}, pulse width, and flux density 
in Table 3.  The spectral index of the mean flux density for the pulse
shown in Fig.~\ref{fig:pfreq} is  $\sim$ -0.7. J2325$-$0530 was not
detected in the periodicity searches that were folded at the period from 
the GBT observations and carried out on each of the detection
observations, despite its high pulse rate and S/N.  A period 
of 0.869 s was found, however, by determining
the least common multiple value between detected pulses. This agrees with 
the period found from timing observations given in~\cite{kar15}.

\begin{table}
\begin{center}
Table 3. Pulse Properties for J2325$-$0530 \\
\begin{tabular}{lllllll}
\hline 
\hline
\vspace{-0.1cm}\\
Frequency & FWHM & Peak Flux & Mean Flux & S/N \\
 (MHz) & (ms) & (Jy) & (Jy) & \\
\hline 
\hline
35.1 & 23.1 $\pm$ 2.2 & 150 $\pm$ 75 & 3.9 $\pm$ 1.9 & 8.8 \\
49.8 & 12.8 $\pm$ 0.9 & 250 $\pm$ 125 & 3.8 $\pm$ 1.9 & 14.1 \\
64.5 & 7.8 $\pm$ 0.6 & 300 $\pm$ 150 & 2.8 $\pm$ 1.4 & 13.2 \\
79.2 & 6.8 $\pm$ 0.9 & 240 $\pm$ 120 & 1.9 $\pm$ 1.0 & 9.8 \\
\hline
\\
\end{tabular}
\end{center}
\end{table}

\section{Discussion}

The detection of only 2 out of 19 RRATs observed is likely a result 
of the low observing frequency of the LWA1 station, which makes
detections difficult due to the effects of pulse broadening by
interstellar scattering, in addition to the spectral properties and burst rates of the targets, and the limited sensitivity of the LWA1.  We note
that the faintest detectable pulses for the LWA1 are at the level
of $\sim$50 Jy compared to $\sim$0.2 Jy for the GBT at 350 MHz
\citep{kar15}.
We do not believe that RFI was an issue, especially
in the 2014 -- 2015 observing campaign, where we carried out the
observations at { night-time} and most runs were almost entirely devoid of
interference other than some short duration burst-like signals across
all DMs.  The rare unusable observations in the second run were simply
re-observed as the RFI was not persistent.  One observation of
J0957$-$06 in the first run tentatively detected a pulse at nearly the
correct DM, but the nature of the pulse could not be confirmed.  The
two observations of this RRAT in the second run did not detect any
pulses at the DM of the pulse in the first run.

\begin{figure}
\begin{center}
\includegraphics[width=3.5in,angle=0]{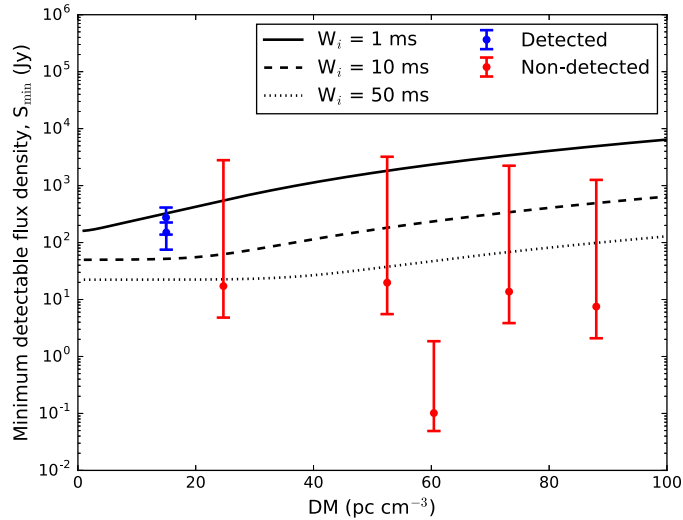}
\end{center}
\vspace{-0.5cm}
\caption{\label{fig:detections} 
Peak flux densities at 57 MHz for the two detected RRATs in our sample
(blue points) and estimated flux densities for non-detections (red
points) calculated using published mean peak flux densities and
assuming a spectral index of $-$1.4. The error bars shown
for each non-detection represent the range of estimated flux densities
corresponding to assumed spectral indices of $-1$ to $-3$. Also shown
are the sensitivity curves of the LWA1 against DM for pulse widths of
1, 10, and 50 ms with the effects of scattering estimated as in
Cordes \& Lazio (2002).
}
\end{figure}

Reasons behind the non-detections can be inferred by investigating the
properties of the RRATs that were observed.  Both of the detected
RRATs have DMs less than 20 pc cm$^{-3}$ with somewhat low detected
pulse rates compared to their pulse rates in the GBT detections.  This
suggests that, not surprisingly, RRATs with DMs greater than 20 pc
cm$^{-3}$ (which make up 14 of the 19 sources observed) 
will be difficult to detect at these low frequencies due to
scattering by the ISM. 
The non-detections may also be explained by RRATs having flatter 
spectral indices, thus reducing the flux densities, and therefore 
observed burst rates, at these frequencies.
The degredation in sensitivity with DM can be seen in 
Fig.~\ref{fig:detections} in which we plot the detected flux density at 57 MHz,
or the range of expected flux densities.  To predict the range in flux density
we assumed a spectral index of between $-$1 and $-$3. 
Only five RRATs had reported flux densities
from higher frequency observations. 
Over this we have plotted the sensitivity of LWA1 to a pulse with a width
of 1, 10 or 50 msec assuming a system equivalent flux density (SEFD) 
for the LWA1 of 9100 Jy, although in practice due to elevation effects
the SEFD for our targets ranged between 6000 and 16000 Jy.  
From this plot we can see that our two detections are both above the
10 msec sensitivity of the LWA1.  { The predicted fluxes for the five
non-detections for which flux measurements are available at high
frequencies, however, fall below the LWA1's threshold, assuming a
spectral index of $-$1.4 (Bates et al. 2013) and scattering as predicted by the NE2001 model (Cordes \& Lazio 2002). The spectral index must be
much steeper, or scattering less significant, in order for these 
sources to be detectable with the LWA1.}

The idea of RRATs being too dim to be observed with the LWA is
somewhat supported by the low DMs of those that we actually detected,
although DM is also correlated with scattering which is perhaps the
more likely culprit.  While pulsars of DMs up to 50 pc cm$^{-3}$ have
been detected with the LWA \citep{sto15}, these were with periodicity
searches where the sensitivity is boosted through the folding of the
data; no such sensitivity advantage exists for single pulse searches.
With that said, none of the RRATs that have been observed in
periodicity searches were detected either, including J1850+15 and
J0628+09, both of which are now also classified as regular pulsars
based on periodicity searches at higher frequencies. J0628+09 has a
very high DM of 88 $\pm$ 2 pc cm$^{-3}$, and a predicted scattering
timescale of 824 milliseconds \citep{NE2001} making detection of
either single or regular pulses with LWA1 unlikely.  Although the DM
of J1850+15 is less than 10 pc cm$^{-3}$ higher than our detected RRATs,
the steep dependence of scattering on DM \citep{bhat04} implies a
scattering timescale four times larger, suggesting that scattering
could have prevented our detection of this pulsar.


{ With few studies having been done at frequencies below 200 MHz,
it is difficult to make conclusions about the intrinsic properties of
RRATs at these low frequencies. However, there have recently been
studies of normal pulsars at low frequencies, which provide a useful
comparison. Normal pulsars often have spectral turnovers around 100 to
200 MHz and also have significant profile evolution.  In some
cases the profile evolution is intrinsic while in others it is due to
scattering~\citep[e.g.][]{sto15}.} It may be that some { RRATs} have a
spectral turnover or a flatter than average spectrum, but it is also
possible to attribute the non-detections to scattering in the
ISM, given average spectral indices.  Additional observations of low
DM RRATs could help to see if some RRATs enter more active phases.
Also the addition of the LWA-SV station and VLA antennas would triple
the sensitivity of the LWA.  Indeed, the largest difficulty currently
in this study is characterized by the high errors on the measurements
of the flux densities, pulse widths, and consequently the spectral
indices.  Improving upon both the accuracy of the fitting and the
sensitivity would increase the value of these measurements. 


\section{Conclusions}

We have observed 19 known RRATs at between 30 MHz and 88 MHz with
LWA1, substantially lower in frequency than their discovery
observations.  Of the targets observed and analyzed in both single
pulse and periodicity search routines, only 2 RRATs were detected and the
pulse widths and flux densities were estimated for each.
We calculated the spectral index of J2325$-$0530 within
the LWA band resulting in a value of $\sim$$-$0.7, which is 
somewhat flatter than for measurements of typical pulsars made at higher 
frequency \citep{Bates13}, possibly indicating there may be a spectral
turnover in J2325$-$0530 as is seen in many normal pulsars. 
We note that these measurements are
taken from a small number of bright pulses, however they are in
agreement with the spectral index distribution of pulsars 
reported by \cite{sto15}, which showed a spectral index 
distribution with a mean of $-$0.7 and a standard deviation of
1.0.  Both of these results
indicate that for those RRATs that the LWA1 can detect, they are much
brighter at these low frequencies.  While the origin of the RRAT's
transient pulsing nature cannot be significantly constrained by this
study, we have demonstrated that RRATs do indeed continue to behave as
transients in this little studied frequency band and that their
pulse profiles are similar to those observed at higher frequencies. 
Both of the detected RRATs had DMs below 20 pc cm$^{-3}$ while 14 of 
the 19 RRATs observed but not detected had DMs above this value,
confirming the notion that scatter broadening of the pulses is an important 
factor at long wavelengths.
Future
studies of RRATs with the LWA or other low frequency instruments will
likely provide further clues into RRATs, their emission mechanisms,
and their relation to the many other subclasses of pulsars.

\section*{Acknowledgements}
The LWA1 is a
University Radio Observatory that is operated by the University of New
Mexico.  Construction of the
LWA has been supported by the Office of Naval Research under Contract
N00014-07-C-0147. Support for operations and continuing development of
the LWA1 is provided by the National Science Foundation under grant
AST-1139974 of the University Radio Observatory program.  
MAM and RM acknowledge support from the National Science Foundation
under grant AST-1327526.  We thank Jasson Hessels and an anonymous
referee for constructive comments.







\label{lastpage}
\end{document}